\documentclass[twocolumn,prl,aps,floats,showpacs,amssymb]{revtex4}
\usepackage{graphicx}
\usepackage{bm}

\begin{document}

\title{Non-universal suppression of the excitation gap in chaotic Andreev
billiards:\\
Superconducting terminals as sensitive probes for scarred states}

\author{Andor Korm\'anyos}
\author{Henning Schomerus}
\affiliation{ Department of Physics, Lancaster University,
Lancaster, LA1 4YB, UK}

\date{May 2006}
\begin{abstract}
When a quantum-chaotic normal conductor is coupled to a
superconductor, random-matrix theory predicts that a gap opens up
in the excitation spectrum which is of universal size $E_g^{\rm
RMT}\approx 0.3 \hbar/t_D$, where $t_D$ is the mean scattering
time between Andreev reflections. We show that a scarred state of
long lifetime $t_S\gg t_D$ suppresses the excitation gap over a
window $\Delta E\approx 2 E_g^{\rm RMT}$ which can be much larger
than the narrow resonance width $\Gamma_S=\hbar/t_S$ of the scar
in the normal system. The minimal value of the excitation gap
within this window is given by $\Gamma_S/2\ll E_g^{\rm RMT}$.
Hence the scarred state can be detected over a much larger energy
range than it is the case when the superconducting terminal is
replaced by a normal one.
\end{abstract}
\pacs{05.45.Mt, 03.65.Sq,  73.21.-b, 74.45.+c}

\maketitle

Typical wavefunctions in classically chaotic quantum systems
\cite{haake,stoeckmann} display a structure-less speckle pattern
which is well captured by Berry's random plain-wave model
\cite{berry}. These quantum systems obey a strong degree of
universality which makes them amenable to a random-matrix theory
(RMT) description. Recent advances of semiclassical techniques
\cite{richtersieber,haakenew,brouwer,jacquod} link the
random-matrix universality to the condition of well-developed wave
chaos. Two sources of deviations from this condition have been
identified. The first source is rooted in the quasi-deterministic
quantum-to-classical correspondence of the dynamics for times
shorter than the Ehrenfest time $t_E$
\cite{richtersieber,haakenew,brouwer,jacquod,ehrenfestreview}.
 The second source is the scarring of
wavefunctions along the trajectories of periodic orbits
\cite{heller,bogomolny,berry2,kaplan,proofs1}.

The complementary nature of both sources of non-universality
becomes apparent when the system is opened up, so that typical
states acquire a width of order $E_{\rm Th}= \hbar/t_D$ (the
Thouless energy), where $t_D$ is the classical mean dwell time in
the system \cite{fyodorov}. Quantum-to-classical correspondence
then induces anomalously short-lived states with resonance width
much larger than $E_{\rm Th}$ \cite{schomerus2004}. Scarred
states, on the other hand, can acquire lifetimes $t_S$ much larger
than the typical lifetime $t_D$ when the openings are not visited
by the periodic orbits which support the scar.
 The resulting narrow resonances of width
$\Gamma_S=\hbar/t_S$ have been observed experimentally and
numerically in a large variety of physical systems, such as
lateral quantum dots \cite{qdots},
microwave cavities \cite{microwaves}, resonant tunnel diodes
\cite{rtd},  micro-optical lasers \cite{microlasers}, and surface
waves \cite{surfacewaves}.

In general, scarred states are a rare fraction of all states of a
system \cite{heller,bogomolny,berry2,kaplan,proofs1} and their
influence can be ignored for almost all energies, with the
exception of the resonance intervals of size $\Gamma_S$. The
purposes of this paper is to demonstrate that scarred states can
be detected over much larger energy ranges when the normal opening
is replaced by a superconducting terminal so that the system forms
a so-called Andreev billiard \cite{Kos}. Chaotic Andreev billiards
have attracted much attention over the past years
\cite{ref:beenakker-review} because they display a gap $E_g$ in
the excitation spectrum which is induced by the superconductor due
to  the dynamical process of Andreev reflection (retro-reflections
at the superconducting interface
 which convert electrons into holes or vice
versa, at the expense of Cooper pairs which are absorbed or
emitted from the superconductor). Random-matrix theory predicts
that this gap is of size $E_g^{\rm RMT}\approx 0.3\,E_{\rm Th}$,
where the Thouless energy refers to the system with the normal
opening (the associated dwell time $t_D$ is identical to the mean
time between successive Andreev reflections in the Andreev
billiard) \cite{ref:beenakker-review,footnote}.

\begin{figure}
\includegraphics[width=\columnwidth]{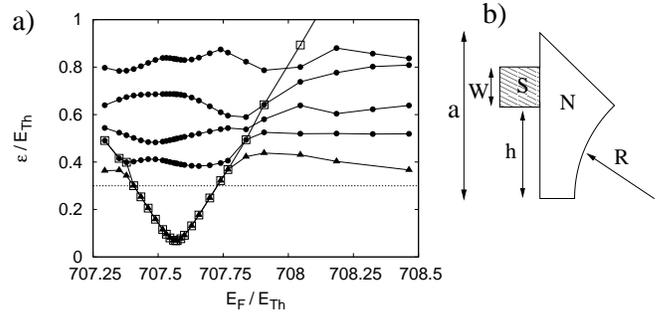}
\caption{Left panel: Excitation spectrum as function of the Fermi
energy $E_F$,  obtained by numerical computations for the Andreev
billiard shown in the right panel. The lowest excitation energy
(triangles) determines the excitation gap. The second, third,
forth and fifth excitation energy are shown as filled circles. The
energy of the scarred eigenstate is highlighted by  open
squares. The dashed line shows the random-matrix prediction
$E_g^{\rm RMT}$ for the gap. Other lines are guides for the eye.
Right panel: The Andreev billiard is a segment of a Sinai billiard with
$W=0.21\,a$, $h=0.58\,a$, and $R=0.8\,a$. The pair potential in
the superconductor is $\Delta_0=0.015 \, E_F$. In the energy range
of the left panel, there are $N=25$ open channels in the
superconducting lead.
\label{fig:geom_and_spectr}}
\vspace*{-5mm}
\end{figure}

We  show that a long-lived scarred state with lifetime $t_S\gg
t_D$ reduces the excitation gap over an energy window $\Delta
E\approx 2 E_g^{\rm RMT}\gg\Gamma_S$. The scar hence determines
fundamental system properties over a much larger range of energies
than in the open normal system. The smallest value of the
excitation gap within the window is given by $E_{g,\,\rm
min}\approx\Gamma_S/2$. The minimal size of the gap is therefore
much smaller than the RMT prediction $E_g^{\rm RMT}$. In
particular, the reduction of the excitation gap can be far
stronger than the reduction due to both the quantum-to-classical
correspondence for times smaller than the Ehrenfest time, observed
in Refs.\ \cite{ref:jacquod,ref:sinai-andreev-rmt}, and the
mesoscopic fluctuations, studied in \cite{ref:vavilov}.

We first present numerical evidence for these results and then
develop a theoretical description which relates the excitation
energy of scarred states in the Andreev billiard to properties of
the scar in the normal billiard.

The excitation energies $\varepsilon$ of an Andreev billiard are
 given by the eigenvalues of the Bogoliubov-de Gennes (BdG) equation
\begin{equation}
\left(\begin{array}[c]{cc}
\hat{{\mathcal H}}_{0}({\bf r}) & \Delta ({\bf r}) \\
\Delta ^*({\bf r}) & -\hat{{\mathcal H}}_{0}^*({\bf r})
\end{array}
\right)
\left(\begin{array}{c}
u ({\bf r})\\ v({\bf r})
 \end{array}\right)=
\varepsilon
\left(\begin{array}{c}
u ({\bf r})\\ v ({\bf r})
\end{array}\right).
\label{eq:BdG}
\end{equation}
Here $\hat{\mathcal{H}}_0=-\frac{\hbar^2}{2m} \nabla^2-E_F$ is the
single-particle Hamiltonian with the chemical potential in the
superconductor set equal to the Fermi energy $E_F$ in the normal
part of the system. The wavefunction is composed of an electronic
part $u(\mathbf{r})$ and a hole part $v(\mathbf{r})$ which are
coupled via the superconducting pair potential
$\Delta(\mathbf{r})$ with value $\Delta_0$ in the bulk of the
superconductor. Typical Andreev billiards satisfy the separation
of energy scales $E_{\rm Th}\ll\Delta_0\ll E_F$. One  can  then
assume that $\Delta(\mathbf{r})=\Delta_0$ is constant in the
superconducting region and vanishes in the normal part of the
system \cite{ref:beenakker-review}. Solutions of Eq.\
(\ref{eq:BdG}) come in pairs of excitation energies $\pm
\varepsilon$, where one solution is of the form ${u\choose v}$ and
the other solution is
 of the form ${-v^*\choose u^*}$.

\begin{figure}
\includegraphics[scale=0.4]{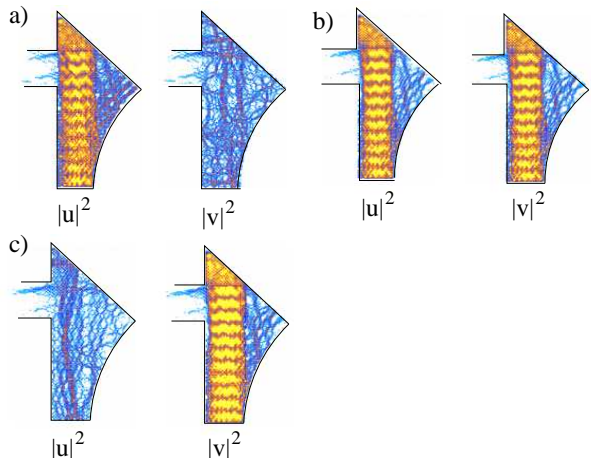}
\caption{ The square moduli $|u|^2$ of the electron and $|v|^2$ of
  the hole components of the wavefunction for  $E_F/ E_{\rm Th}=
  707.434 < E_{F,\rm min}$ (a), for $E_F= E_{F,\rm min}$ (b) and for
 $E_F/ E_{\rm Th}= 707.695 > E_{F,\rm min}$ (c).
\label{fig:scarred-states}}
\vspace*{-3mm}
\end{figure}

 In order to convince ourselves of the visibility of scars in
the excitation spectrum of an Andreev billiard we solved the
Bogoliubov-de Gennes equation numerically for the chaotic billiard
 shown in the right
panel of Fig.~\ref{fig:geom_and_spectr}
\cite{ref:sinai-andreev-rmt,ref:SA-wavefunc}. The normal part is a
de-symmetrized segment of a Sinai billiard. The superconducting
terminal is attached to one of the straight walls. The left panel
of Fig.~\ref{fig:geom_and_spectr} shows the positive excitation
energies as a function of the Fermi energy $E_F$, scaled in units
of the Thouless energy $E_{\rm Th}$ (the dependence of the
Thouless energy on the Fermi energy is taken into account in this
rescaling). The second, third etc. excitation energies show the
characteristic level dynamics of chaotic systems when one
parameter is changing, and avoided crossings can also be observed.
Random matrix theory predicts that the excitation gap (set by the
the lowest positive excitation energy) fluctuates around the value
$E_{g}^{\rm RMT}=0.3 E_{\rm Th}$ (dashed line)
\cite{ref:beenakker-review,footnote}. However, there is an energy
interval of width $\Delta E/ E_{\rm Th}\approx 0.26 $ in which the
excitation gap drops distinctively below this value. The smallest
gap $E_{g, \rm min} =0.0654 E_{\rm Th}$ is attained at Fermi
energy $E_{F,\rm  min} = 707.567 E_{\rm Th}$.

Inspection of the wavefunction of the lowest eigenstate in this
energy interval reveals an interesting feature: for $E_F$ smaller
than $E_{F, \rm min}$ [see Fig.~\ref{fig:scarred-states}(a)], the
electronic component $u$ is nonuniform and shows clear signatures
of scarring along a family of marginally stable periodic orbits
which bounce between the straight segments of the billiard.
Calculating the probability $P_e=\int |u|^2\,d{\bf r}$ ($P_h=\int
|v|^2\,d{\bf r}$) of finding the quasiparticle in the electron
(hole) state in the normal part of the system, one finds
$P_e=0.915$ and $P_h=0.078$, {\em i.e.} the quasiparticle is
dominantly electron-like (the remaining weight is in the
superconductor). The next panel [Fig.~\ref{fig:scarred-states}(b)]
corresponds to the eigenstate at the minimum of the excitation gap
in Fig.~\ref{fig:geom_and_spectr}(a), showing that in this case
both $u$ and $v$ are nonuniform, and the probabilities are
$P_e=0.520$ and $P_h=0.474$. In Fig.~\ref{fig:scarred-states}(c)
we show an eigenstate at a  Fermi energy  which is larger than
$E_{F,\rm min}$. Here the hole component is scarred by the same
family of periodic orbits and the quasiparticle becomes dominantly
hole-like with $P_e=0.064$ and $P_h=0.929$.

\begin{figure}[htb]
\includegraphics[scale=0.5]{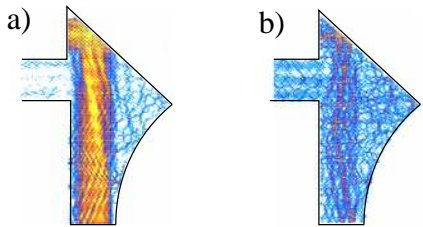}
\caption{Scattering wavefunction of the open normal dot at
energies $E=E_{F,\rm  min}$ (a) and  $E=707.657 E_{\rm Th}$ (b).
The states are excited by an incoming wave in the third transverse
mode of the lead. \label{fig:scat-wavefunc}}
\end{figure}

 These observations  suggest that the suppression of the excitation gap
and the scarring of the wavefunction  stems from a long-lived
scarred quasibound state of the normal open billiard. When the
superconducting terminal is replaced by a normal lead, a scarred
state can indeed be  detected for energies very close to $E_{
F,\rm min}$ -- see the left panel of Fig.~\ref{fig:scat-wavefunc},
which shows a scattering state computed at the energy $E=E_{F,\rm
min}$. For a slightly different energy $E=707.657 E_{\rm Th}$,
however, this scar is already very much diminished, as is shown in
the right panel of Fig.\ \ref{fig:scat-wavefunc}. At this energy,
the gap in the excitation spectrum of the Andreev billiard is
still strongly suppressed. These numerical results demonstrate
that a long-lived scarred state can be detected over an energy
window much larger than its resonance width $\Gamma_S$ in the
normal billiard when the system is coupled to a superconducting
terminal.

 We now present
a theory which explains the numerical observations. In particular,
we relate the Fermi-energy dependence of the scarred excitation in
the Andreev billiard to the properties of the scarred state in the
open normal billiard. Such relations can be derived because the
excitation spectrum of  Andreev billiards can be calculated from
the scattering matrix $S(E)$ of the normal system using the
quantization condition \cite{ref:beenakker-review}
\begin{equation}
\det [1+S^*(E_F-\varepsilon) S(E_F+\varepsilon)]=0
\label{eq:quant-cond} ,
\end{equation}
which is valid in the limit $\varepsilon\ll \Delta_0 \ll E_F$. The
scattering matrix is a unitary (and in absence of a magnetic field
also symmetric) matrix of dimensions $N\times N$, given by the
number of open channels $N$ in the opening. In
Eq.~(\ref{eq:quant-cond}) $S(E_F+\varepsilon)$
[$S^*(E_F-\varepsilon)$] describes the propagation of  electrons
[holes] and $*$ denotes complex conjugation. Equation
(\ref{eq:quant-cond}) embodies the condition of total constructive
interference after two Andreev reflections, so that an electron is
converted first to a hole and then back to an electron. (Each
Andreev reflection contributes a phase factor of $i$.)

Narrow resonances in the normal system are associated to poles of
the scattering matrix which are situated close to the real axis in
the lower half of the complex energy plane. In what follows we
will assume that there is only one such narrow resonance in the
energy range of interest, namely that  associated to the scarred
state, located at a complex energy  ${\cal E}=E_S-i\, \Gamma_S/2$.
In order to isolate the contribution of this narrow resonance to
the scattering matrix  we first diagonalize  $S(E)=O(E)\Lambda(E)
O^T(E)$ where $O(E)$ is an $N\times N$ dimensional orthogonal
matrix and $\Lambda(E)$ is a diagonal matrix containing the
eigenvalues $\lambda_i(E)$, $i=1\dots N$.
 In the vicinity of the narrow resonance the energy
dependence of $O(E)$ can  be neglected\cite{ref:noeckel}. Inserting  the
decomposition $S(E)=O \Lambda(E) O^T$ into
Eq.~(\ref{eq:quant-cond}) one then finds that the quantization
condition can be written as
\begin{equation}
\prod_i[1+\lambda^*_i(E_{F}-\varepsilon)\lambda_i(E_{F}+\varepsilon)]=0.
\label{eq:qcond2}
\end{equation}
One of the eigenvalues,  $\lambda_S(E)$, is associated to the
narrow resonance, while the other eigenvalues are associated to
the remaining  non-resonant states in the system. The resonant
eigenvalue   can be  approximated  as
$\lambda_S(E)=e^{2i\theta_S(E)}\frac{E-{\cal E}^*}{E-{\cal E}}$
\cite{ref:noeckel} where the energy dependence of $\theta_S(E)$
can again be neglected over the energy range of interest.
Demanding that  the term involving $\lambda_S$ in
Eq.~(\ref{eq:qcond2}) be zero,
\begin{equation}
1+\frac{E_{F}-\varepsilon_S-E_S-i\,\Gamma_S/2}{E_{F}-\varepsilon_S-E_S+i\,\Gamma_S/2}
\frac{E_{F}+\varepsilon_S-E_S+i\,\Gamma_S/2}{E_{
F}+\varepsilon_S-E_S-i\,\Gamma_S/2}=0,
\end{equation}
one can find  the excitation energies
\begin{equation}
\varepsilon_S=\pm\sqrt{(E_F-E_{S})^2+\Gamma_{S}^2/4 }
\label{eq:result}
\end{equation}
associated to the \emph{scarred eigenstate of the Andreev
billiard} in terms of the energy and width of the \emph{scarred
resonance in the open normal system}. The excitation energies come
in pair as required by the particle-hole symmetry of the BdG
equation (\ref{eq:BdG}).

For Fermi energies far less than the energy $E_S$ of the scar in
the normal system, the positive excitation energy is given by
$\varepsilon_S \approx E_S-E_{F}$, which is then  far larger than
the RMT gap and hence corresponds to a highly excited state. As
$E_{F}$ is increased, the excitation drifts down through the
spectrum, and assuming $\Gamma_S \ll E_g^{\rm RMT}$ the scarred
state eventually becomes the lowest excited state at an energy
$E_F\approx E_S-E_{g}^{\rm RMT}$.
 This
defines the beginning of the energy interval in which the scar can
be detected by measuring the excitation gap of the Andreev
billiard. The minimal value $ E_{g,\rm
min}=\min_{E_F}|\varepsilon_{S}|=\Gamma_S/2$ of the excitation gap
is obtained for $E_{F, \rm  min}=E_S$. When the Fermi energy is
further increased the excitation energy increases as well, and for
values $E_F\gtrsim E_S+E_{g}^{\rm RMT}$ the scarred state becomes
again a higher excited state, with energy $\varepsilon_S \approx
E_F-E_{S}$.  This explains the qualitative features observed in
Fig.~\ref{fig:geom_and_spectr}(a).

\begin{figure}[tbh]
\includegraphics[width=\columnwidth]{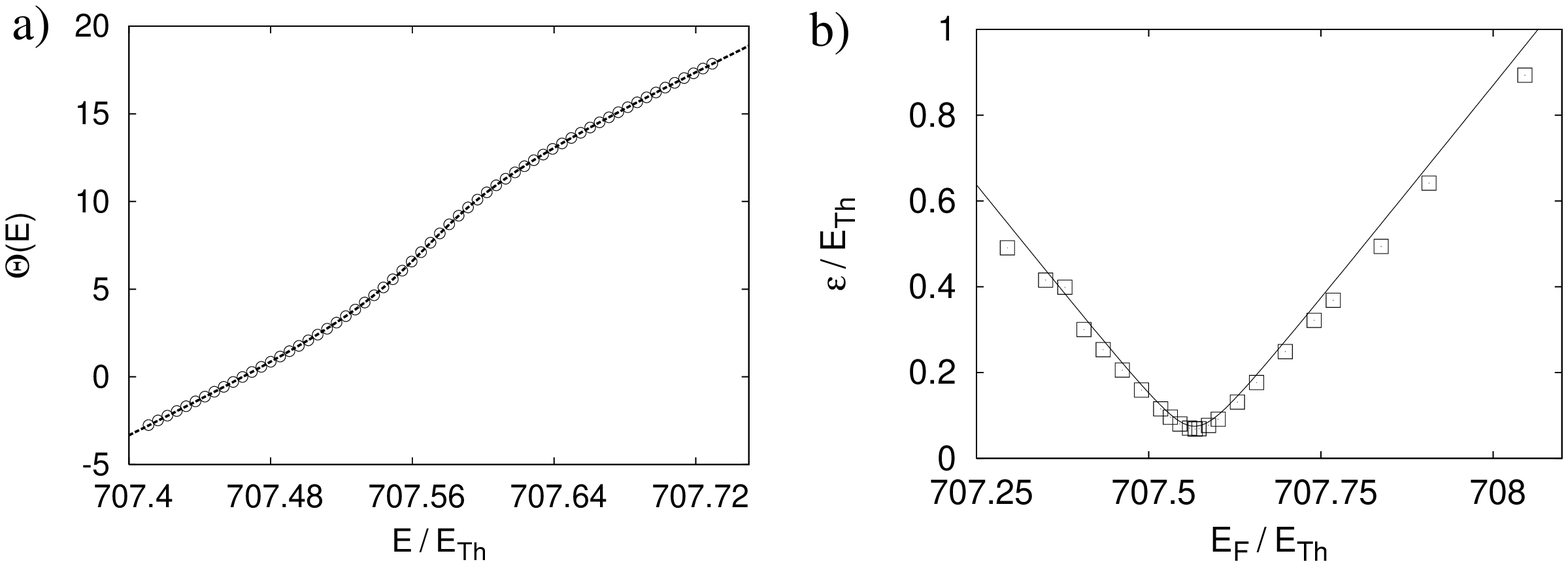}
\caption{Left panel: The phase $\Theta(E)={\rm Im} \ln\det[S(E)]$
of the scattering matrix of the open normal billiard [Fig.\
\ref{fig:geom_and_spectr}(b) with the superconducting terminal
replaced by a normal lead] is plotted as a function of energy
(circles). The solid line shows a fit of the function
$2\arctan\frac{\Gamma_S}{2(E-E_{S})}+C\,E+\Theta_0$ to
$\Theta(E)$. Right panel: Quantitative comparison of the
excitation energies of the scarred Andreev bound state (squares)
with Eq.~(\ref{eq:result}) (solid line), using the parameters
$E_S$ and $\Gamma_S$ of the scar in the normal system.
\label{fig:phase-fit}} \vspace{-5mm}
\end{figure}

The microscopic parameters $E_S$ and $\Gamma_S/2$ entering Eq.\
(\ref{eq:result}) can be determined directly from properties of
the open normal billiard. The left panel of
Fig.~\ref{fig:phase-fit} shows the phase $\Theta(E)={\rm Im}\ln
\det S(E)$ of the scattering matrix over the energy range in which
the scar determines the excitation gap. The energy dependence due
to the resonant eigenvalue $\lambda_S$ is given by ${\rm Im} \ln
\lambda_S\approx 2\, \arctan\frac{\Gamma_S}{2(E-E_S)}$, while the
non-resonant states in the system contribute a linear background
$CE+\Theta_0$ with $C\approx 2\pi/\delta_N$, where $\delta_N$ is
the mean level spacing of the normal system. A fit  of the curve
$2\arctan \frac{\Gamma_S}{2(E -E_S)}+CE+\Theta_0$ to the numerical
data delivers the values $E_S= 707.567 E_{\rm Th}$ and
$\Gamma_{S}/2 = 0.0745 E_{\rm Th}$. The width is in good agreement
with the smallest value $E_{g,\rm min}$ of the excitation gap
given above, and the energy of the scar coincides exactly with the
Fermi energy $E_{F,\rm min}$ at which this minimal value is
attained. The right  panel of Fig.~\ref{fig:phase-fit} shows that
substituting the obtained parameters $E_S$ and $\Gamma_S$ into
Eq.~(\ref{eq:result}), a \emph{quantitative} agreement can be
found between the theoretical prediction and the computed values
of the excitation energies of the scarred states.

Finally, we turn to the observation that the scarred character of
the state in the Andreev billiard shifts from the electronic
component for $E_F\ll E_{F, \rm  min}$ to the hole component for
$E_F\gg E_{F,\rm min}$, while both components show an equal amount
of scarring for $E_F=E_{F,\rm min}$ (see Fig.\
\ref{fig:scarred-states}).
 The weight of the
scar in each component can be defined by the squared overlaps
$P_{e,S}=|\int\, \psi_S^* u \,d{\bf r}|^2$ and  $P_{h,S}=|\int\,
\psi_S^* v \,d{\bf r}|^2$ with the scarred wavefunction $\psi_S$
in the normal billiard. To a good approximation, these weights
should be proportional to the time spent as a particle or hole
excitation, which can be quantified by the Wigner delay time $
\tau_W(E)=\hbar \frac{d \Theta}{d E}$ \cite{fyodorov}. Within our
previous approximations, we can write
\begin{equation}
\tau_W(E) =
\frac{\hbar\Gamma_{S}}{(E-E_{S})^2+\Gamma_{S}^2/4}+\hbar C.
\end{equation}
Ignoring the constant $C$ which accounts for the background of
non-resonant states we find
\begin{equation}
\frac{P_{e,S}}{P_{h,S}}\approx
\frac{(E_F-\varepsilon_S-E_{S})^2+\Gamma_{S}^2/4}{(E_F+\varepsilon_S-E_{S})^2+\Gamma_{S}^2/4}
= \frac{\varepsilon_S-(E_F-E_{S})}{\varepsilon_S+(E_F-E_{S})} .
\end{equation}
For $E_F\ll E_S$ this delivers $P_{e,S}\gg P_{h,S}$, hence
scarring in the electronic part of the wavefunction, while the
hole part is scarred for $E_F\gg E_S$. For $E_F=E_S$ the weights
in both components are predicted to be equal. This explains the
observations in Fig.\ \ref{fig:scarred-states}.

In summary, we have demonstrated  theoretically and numerically
that long-lived scarred wavefunctions of quantum-chaotic system
can be observed over large ranges of the Fermi energy when the
system is connected to a superconducting terminal. The resulting
suppression of the excitation gap should be accessible in
experiments by tunnel spectroscopy \cite{Devoret}. The Fermi
energy can be controlled by side gates \cite{Takayanagi}.

This work was supported by the European Commission, Marie Curie
Excellence Grant MEXT-CT-2005-023778 (Nanoelectrophotonics).


\begin{thebibliography}{99}
\bibitem{footnote}
In this paper we work with the conventional Thouless energy
$E_{\rm Th}=\hbar/t_D$ of the normal open system, which differs by
a factor of two from the Thouless energy $\hbar/2t_D$ often used
for superconducting systems \cite{ref:beenakker-review}.
\end{thebibliography}
\end{document}